# Broadband Optical Fully Differential Operation Based on the Spin-orbit Interaction of Light

Shanshan He[1], Junxiao Zhou[1], Shizhen Chen[1], Weixing Shu[2], Hailu Luo[1*] and Shuangchun Wen[2]

[1]*Laboratory for Spin Photonics, School of Physics and Electronics, Hunan University, Changsha 410082, China*

[2]*Key Laboratory for Micro-/Nano-Optoelectronic Devices of Ministry of Education, School of Physics and Electronics, Hunan University, Changsha 410082, China*

*Hailu Luo* hailuluo@hnu.edu.cn

**Abstract**

**Optical technology may provide important architectures for future computing, such as analog optical computing and image processing. Compared with traditional electric operation, optical operation has shown some unique advantages including faster operating speeds and lower power consumption. Here, we propose an optical full differentiator based on the spin-orbit interaction of light at a simple optical interface. The broadband optical operation is independent on the wavelength due to the nature of purely geometric. As an important application of the fully differential operation, the broadband image processing of edge detection is demonstrated. By adjusting the polarization of the incident beam, the one-dimension edge imaging at any desirable direction can be obtained. The broadband image processing of edge detection provides possible applications in autonomous driving, target recognition, microscopic imaging, and augmented reality.**

**Introduction**

In recent years, compared with the maturity of digital circuit technology, there has been a lot of optimism about the future development of optical analog computing[1,2]. While digital signal processors provide high speed and reliable operation, it also has some obvious disadvantages, such as high-power consumption, expensive analog-to-digital converters, and sharp performance degradation at high frequencies[3]. Considering these limitations, it is neither rational nor economical to use digital signal processors to perform specific, simple computing tasks such as differential or integral, equation solving, matrix inversion, edge detection, and image processing[4-11]. Analog signal processors due to their wave-based characteristics potentially have superior performance over digital versions, including faster operating speeds and lower power consumption[12-18]. Based on the above advantages, the differentiator is used to realize ultra-fast parallel image processing, real-time boundary detection and judgment of boundary direction[19,20]. It has important applications in high speed, high throughput image processing (satellite and medical image)[21-25], parallel computing[26-36], and is expected to be used in artificial intelligence such as optical neural network[37], and high speed pattern recognition[21].

With the rapid development of information technology, the demand of information processing performance is constantly improving. Because of the advantages of optical information processing technology, such as superfast, large bandwidth, large flux, and low loss, it has become an important topic in recent years to realize optical information processing by using micro-integrated and micro-nano structure. In the previous reports, the design of micro-nano structure is extremely complex[11,38], so it is very difficult to fabricate these devices accurately and realize optical simulation operation. In response to this challenge, optical differentiators based on surface plasmon[21], photonic spin Hall effect,[25] and metasurface[39-41] have been developed. However, the surface plasma-excited differentiator and the differentiator based on photonic spin Hall effect can only achieve partial differentiation, and the metasurface-based differentiators which are expensive to manufacture have bandwidth limitations. There are challenges in achieving broadband fully differentiation.

In our work, we present a differentiator to realize broadband fully differentiational operation. The differentiator is consisting of two orthogonal polarizers and a glass plate, which can achieve spatial differentiation in the beam reflection process, and carry out the fully differentiation of the input light field. When the polarization of the incident beam is changed, and always ensuring that the polarization of the analyzer is perpendicular to the linear polarization direction on the center wave vector of the reflected beam, we experimentally achieve one-dimensional edge image in any desirable direction. In other words, this full differentiator can control the direction of edge detection by rotating the incident polarization. On the other hand, the differentiator is based on the principle of spin-orbit interaction of light, and we have demonstrated that its working performance has no wavelength limit.

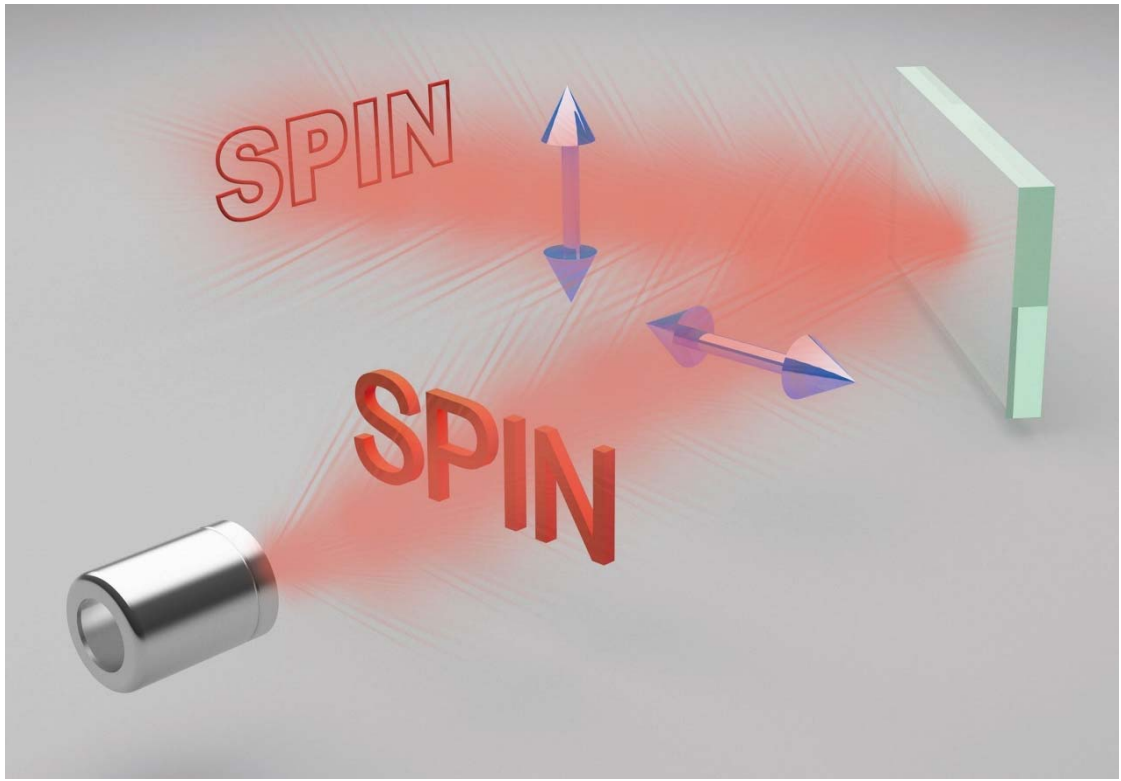

**Fig. 1 Schematic of optical full differentiator based on the spin-orbit interaction of light at an optical interface.** The blue arrows represent the polarization axis of two related polarizers, these two polarizers and the air-glass reflection interface constitute the optical fully differentiator, which is fed a complete "SPIN" image and outputs its edge image.

## Results

### The concept of optical fully differential operation

In order to achieve optical fully differential operation, a schematic illustration of the optical full differentiator is shown in Fig. 1. When the input target image is processed by this differentiator, the output edge image of the target will be obtained. To understand the generalized spatial differentiation, we consider the input photons with arbitrary linear polarization state:

$$|\psi_i\rangle = \cos\gamma_i |H\rangle + \sin\gamma_i |V\rangle, \tag{1}$$

where $\theta_i$ is the incident angle, $\gamma_i$ is the polarization angle. Based on the boundary condition, the polarization state of reflected photons can be written as[42-45]

$$|\psi_r\rangle \approx \exp(+ik_{rx}\Delta x + ik_{ry}\Delta y)\exp(-i\gamma_r)|+\rangle + \exp(-ik_{rx}\Delta x - ik_{ry}\Delta y)\exp(+i\gamma_r)|-\rangle, \tag{2}$$

where $\gamma_r = \arctan\left(\frac{r_s}{r_p}\tan\gamma_i\right)$ is the polarization angle of reflected photons, $k_{rx}$ and $k_{ry}$ are wave vectors components of the reflection beam, $r_p$ and $r_s$ are the Fresnel's reflection coefficients of the p and s polarizations, respectively. Here, we have introduced the approximations: $\exp i\sigma_3(k_{rx}\Delta x + k_{ry}\Delta y) \approx 1 + i\sigma_3(k_{rx}\Delta x + k_{ry}\Delta y)$, where $\sigma_3$ is the Pauli operator. The origin of the spin-orbit interaction terms $\exp i\sigma_3(k_{rx}\Delta x + k_{ry}\Delta y)$ lies in the transverse nature of the photon polarization: The polarizations associated with the plane-wave components experience different rotations in order to satisfy the transversality in reflection[46]. The term of $\Phi_G = k_{rx}\Delta x + k_{ry}\Delta y$ is the geometric phase known as the spin-redirection Berry phases[47,48].

The wave function in position space can be obtained by Fourier transformation for given wave function in momentum space $\varphi_{in}(k_{ix}, k_{iy})$. The total wave function is made up of the packet spatial extent and the polarization state, and is given by

$$|\varphi_r(x,y)\rangle = \varphi_{in}[(x+\Delta x),(y+\Delta y)]\exp(-i\gamma_r)|+\rangle + \varphi_{in}[(x-\Delta x),(y-\Delta y)]\exp(+i\gamma_r)|-\rangle. \tag{3}$$

In the spin basis, the coupling between the spin state and the transverse momentum of light leads to the splitting of the wave-packets correlated with different spin states. The spin-dependent shifts including in-plane spin separation $\Delta x$ and transverse spin separation $\Delta y$ are well-known as photonic spin Hall effect[42,43].

When the reflected photons pass through an analyzer whose polarization axis is orthogonal to the polarization of center wave vector $\gamma_{p2} = \frac{\pi}{2} + \gamma_r$. Hence, the final wave function going through the whole differentiator system can be rewritten as

$$\varphi_{out}(x,y) = \varphi_{in}[(x+\Delta x),(y+\Delta y)] - \varphi_{in}[(x-\Delta x),(y-\Delta y)]. \tag{4}$$

In general, the spin-dependent shifts are much smaller than the profile of input wave function due to the weak spin-orbit interaction of light. Therefore, $\varphi_{out}(x,y)$ can be approximately written as the spatial fully differentiation of the input $\varphi_{in}(x,y)$:

$$\varphi_{out}(x,y) \simeq \Delta x \frac{\partial \varphi_{in}(x,y)}{\partial x} + \Delta y \frac{\partial \varphi_{in}(x,y)}{\partial y}. \tag{5}$$

Here, $\Delta x$ and $\Delta y$ are increments which can be modulated by the incident polarization $\gamma_i$. It should be noted that spatial full differentiation is not limited by the wavelength due to the purely geometric nature of spin-orbit interaction of light. Moreover, the spatial full differentiation can be adopted in the classical level with a large number of photons in a quantum coherent state, where each photon behaves independently and the light is treated coherently in the paraxial regime[44].

**Spatial spectral transfer function**

In order to prove that the optical fully differentiator can achieve the performance we desired, we designed experiments to verify that the splitting direction of the output beam rotates with the change of incident polarization. Further, it is proved concretely by spatial spectral transfer function that the differentiator presents spatial differentiation in different directions by adjusting the incident linear polarization state. The experimental device is shown in Fig. 2, HWP adjusts the polarization of a collimated Gaussian laser beam, with the wavelength of 632.8 nm, to control the intensity of incident beam. The combination of L1 and L2 makes up a 4f system, and here the distance from L1 to L2 is set to the sum of the focal lengths of the two lenses. The bevel of a rectangular prism (N-BK7) provides a reflective surface and the angle of incidence is $\theta_i = 30°$. The polarizer GLP1 is used to prepare the initial field polarization state, GLP2 is used to analyze the reflection field polarization state, and their optical direction satisfies $\gamma_{p2} = \frac{\pi}{2} + \arctan\left(\frac{r_s}{r_p}\tan\gamma_i\right)$. The components of GLP1, GLP2 and prism form the optical fully differentiator, placed in the between of two lenses. The CCD is placed on the Fourier plane (the back focal plane) of L2.

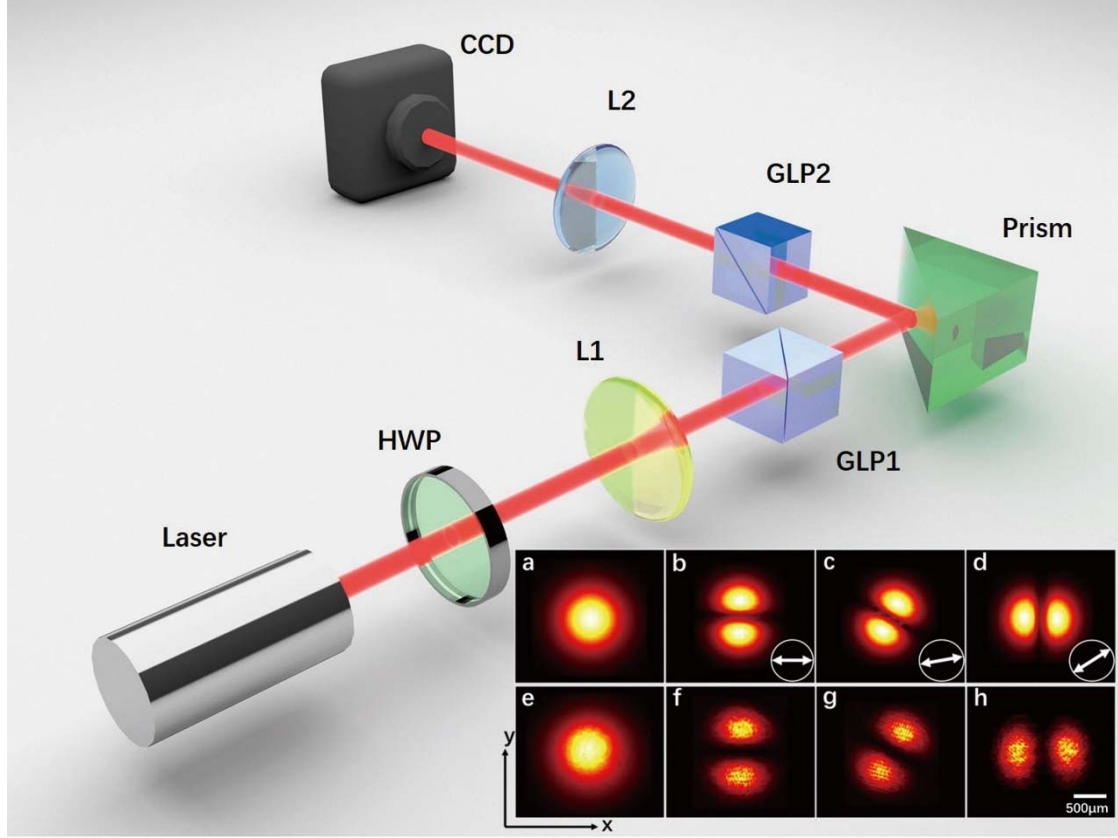

**Fig. 2 Spatial differentiation demonstration for the optical fully differentiator.** The light source is a He-Ne laser (wavelength $\lambda = 632.8$ nm); HWP, half-wave plate; L, lens; GLP, Glan laser polarizer; inclined surface of Prism (N-BK7) acts as the air-glass reflection interface; CCD, charge-coupled device. The insets correspond to spatial differentiation results for a Gaussian illumination with an incident angle $\theta_i = 30°$. **a-d** Theoretical light fields, **e-h** The experimental results. The white double arrow indicates the optical direction ($\gamma_i$) of GLP1. **a, e** Intensity profiles of the incident beam. **b, f** Intensity profiles of the reflected beam with a direction angle of the incident linear polarization $\gamma_i = 0°$. **c, g** Intensity profiles of the reflected beam with $\gamma_i = 10°$. **d, h** Intensity profiles of the reflected beam with $\gamma_i = 40°$.

According to Eq. 2, the spin-dependent shifts ($\Delta x$ and $\Delta y$) vary with the direction of the incident linear polarization (see Supplementary Eq. S11,12). When the incident beam is in horizontal polarization state ($\gamma_i = 0°$) or vertical polarization state ($\gamma_i =$

90°), there is no in-plane spin separation $\Delta x$ but transverse spin separation $\Delta y$; when $\gamma_i$ is equal to 40°, only $\Delta x$ exists; on the other hand, the incident beam is in other linear polarization states, $\Delta x$ and $\Delta y$ exist simultaneously (see Supplementary Fig. S1). We experimentally select three points, severally, $\gamma_i = 0°$, $\gamma_i = 10°$ and $\gamma_i = 40°$, to observe the splitting direction of the output beam. Figures 2a-d show the theoretical intensity profiles for the incident and reflected beams. Figures 2e-h show the measured results. We find that the splintered reflected beams observed in the experiment are consistent with the shape of the theoretical simulation.

To illustrate the optical full differentiation effect, we measure the spatial spectral transfer function under a Gaussian beam illumination. We assume that the wave function in momentum space can be specified by the following expression:

$$\varphi_{in}(k_{ix}, k_{iy}) = \frac{w_0}{\sqrt{2\pi}} exp\left[-\frac{w_0^2(k_{ix}^2 + k_{iy}^2)}{4}\right], \tag{6}$$

where $w_0$ is the width of the wave function. The total wave function is made up of the packet spatial extent and the polarization state.

We further processed the data from Fig. 2 to quantitatively analyze spatial differentiation. Light intensity is proportional to the square of the amplitude of the electric field $I(x, y) \propto |\varphi(x, y)|^2$. Therefore, the field-profile transformation from the incident to the reflected light is described by a spatial spectral transfer function

$$H(k_x, k_y) = \varphi_{out}(k_x, k_y)/\varphi_{in}(k_x, k_y), \tag{7}$$

where $\varphi(k_x, k_y)$ is the Fourier transform of $\varphi(x, y)$. Figure 3 shows the theoretical and experimental transfer function.

Figure 3d is generated by processing data extracted from Fig. 2e and Fig. 2f, corresponding to the spatial differentiation in the *y* direction. Figure 3e is formed by processing data extracted from Fig. 2e and Fig. 2g. It reveals that the spatial differentiation includes two components in the *x* direction and *y* direction. Figure 3f is obtained by processing data extracted from Fig. 2e and Fig. 2h, corresponding to the spatial differentiation in the *x* direction. The measurement results of transfer function also prove that the differentiator can produce two variables ($\Delta x$, $\Delta y$) at the same time and realize full differentiation. And the direction of spatial differentiation depends on the synthetic direction of the two variables.

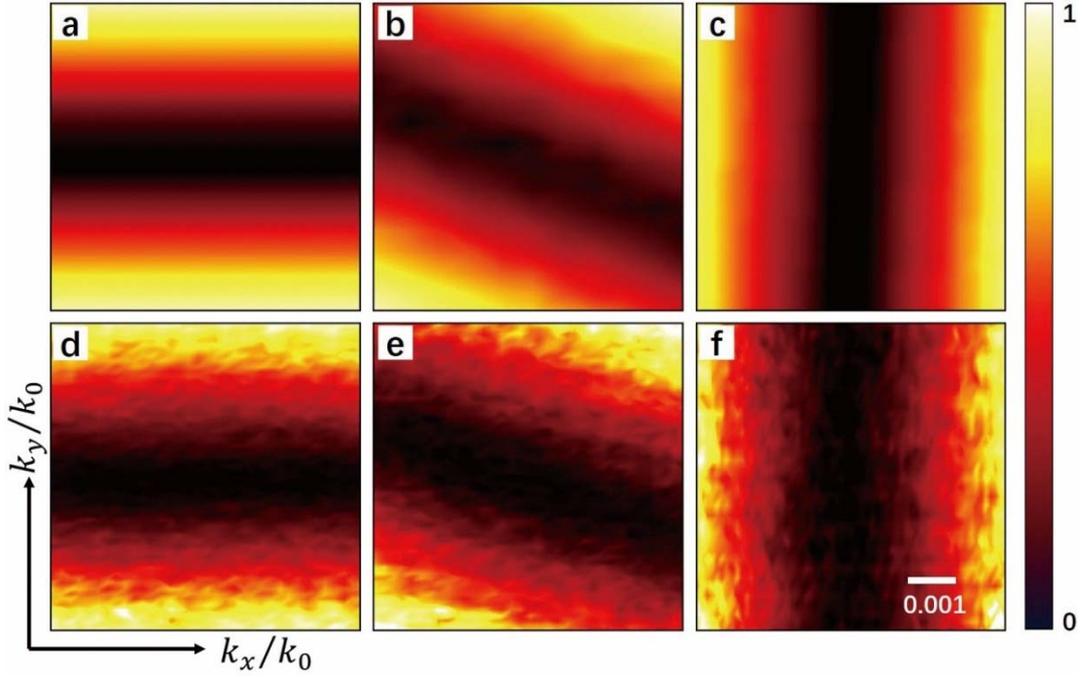

**Fig. 3 Measurement of the spatial spectral transfer function on an air-glass interface. a**, **d** Theoretical simulation and measured spatial transfer function spectrum with a direction angle of the incident linear polarization $\gamma_i = 0°$, respectively. **b**, **e** Theoretical simulation and measured spatial transfer function spectrum with $\gamma_i = 10°$, respectively. **c**, **f** Theoretical simulation and measured spatial transfer function spectrum with $\gamma_i = 40°$, respectively. The incident angle is chosen as $\theta_i = 30°$.

## Image Processing of Edge Detection

We now demonstrate an important application of our optical fully differentiator for the edge image detection. The experimental apparatus is shown in Fig. 4. The USAF-1951 resolution target is used as an object, we select the number 2, the letter S, and horizontal stripes to conduct the experiment of edge detection. The L1 and L2 form the first 4f system, the object is placed at the front focal plane of L1, with the purpose of transferring the target image to the air-glass interface. L3 and L4 constitute the second 4f system, the optical path between L2 and L3 is the sum of their focal lengths, the CCD is placed at the rear focal plane of L4 to transfer the edge image to the CCD induction region. HWP adjusts the polarization of the incident beam, which works the same way as in Fig. 2. The optical full differentiator consisting of GLP1, GLP2, and prism is located between L2 and L3, the incident angle is chosen as $\theta_i = 30°$. The polarization state of the incident light field is modulated by rotating the polarization axis of GLP1, while the intermediate light field of the target image is eliminated by GLP2 and preserves the edge image.

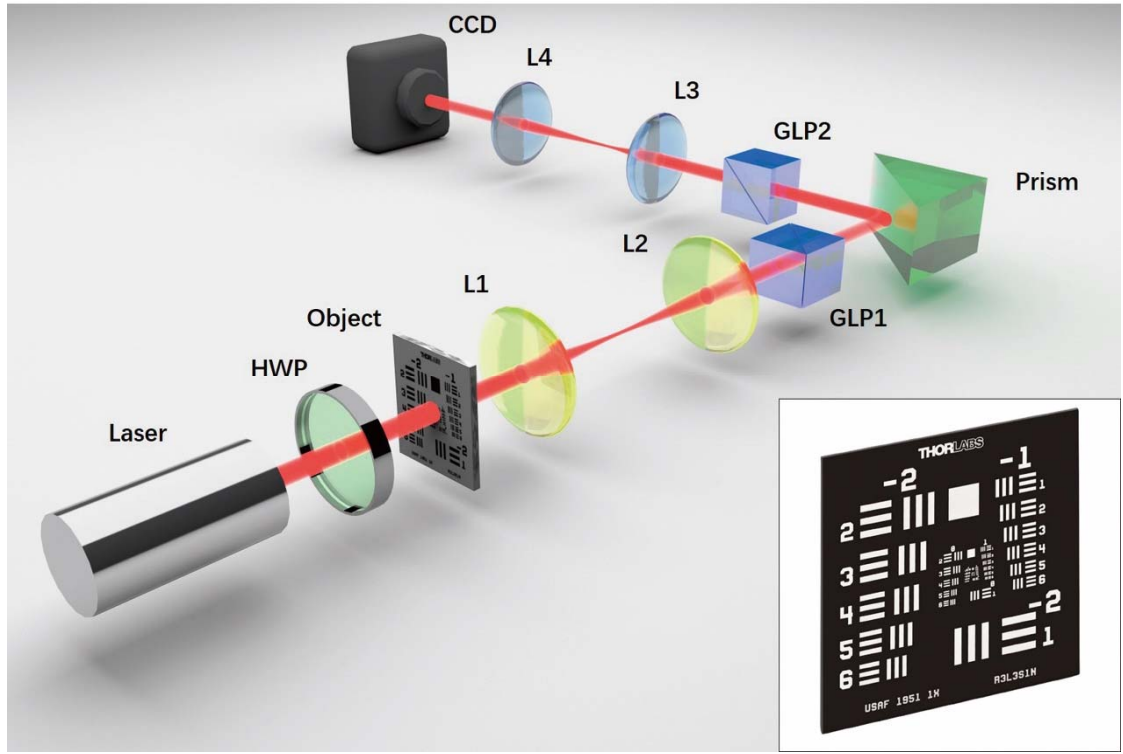

**Fig. 4 Experimental setup for edge detection based on optical fully differentiator.** L1 and L2 form the first 4f system, the object (resolution target) is placed at the front focal plane of L1. L3 and L4 constitute the second 4f system, the CCD is placed at the rear focal plane of L4. The optical path between L2 and L3 is the sum of their focal lengths.

Figure 5 shows edge detection of number 2, the letter S and horizontal stripes images of for He-Ne laser (Thorlab) with wavelength of 632.8nm. First, we obtain the target images which comprise the full information obtained by removing GLP2. Then, we add GLP2 and adjust its polarization axis to dim the light field in the middle of the target image until the CCD captures the corresponding edge detection image. When the incident linear polarization states are controlled (the polarization axis of GLP1 to be 0°, 10° and 40°), the results of edge detection are consistent with the spatial differentiation shown by the spatial spectral transfer function. Therefore, the direction of edge detection can be controlled by adjusting the incident polarization, and the edge state in any desirable direction of the image can be obtained. Furthermore, the complete target edge image can be restored according to the edge images in different directions.

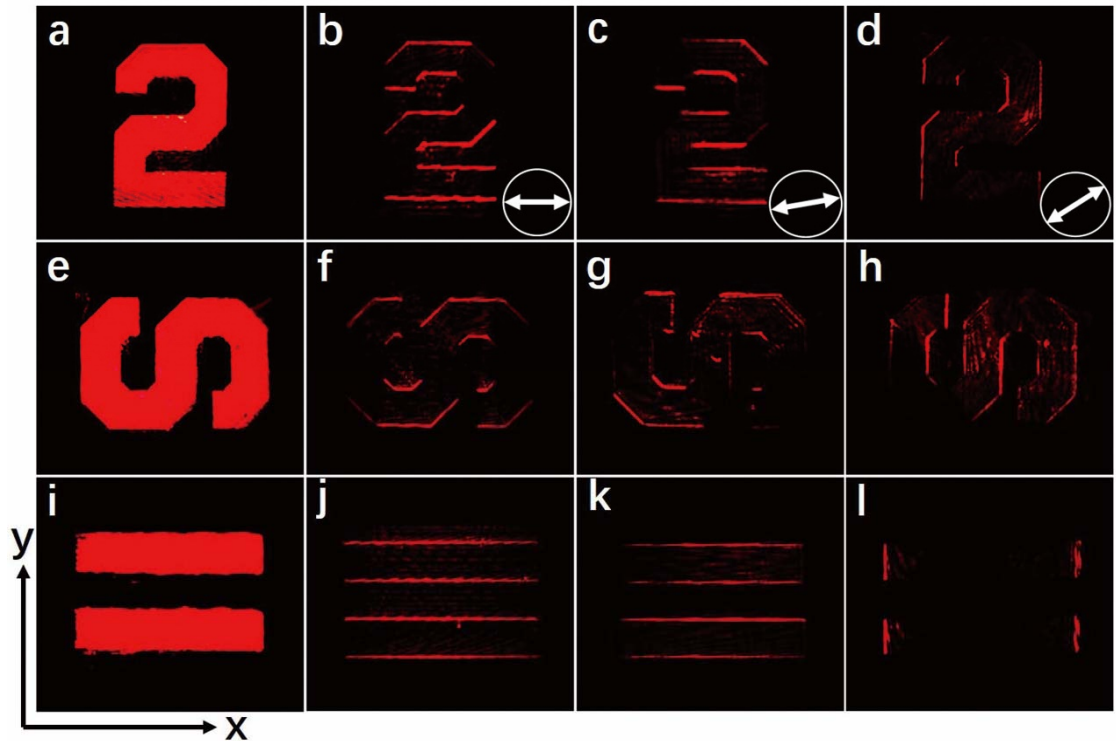

**Fig. 5 Edge detection for He-Ne laser (λ = 632.8 nm) with an incident angle $\theta_i = 30°$. a**, **e** and **i** are different incident images of number 2, the letter S and horizontal stripes, respectively. The white double arrow indicates the polarization direction of incident beam $\gamma_i$. **b**, **f** and **j** are the target edge images of three different patterns with a direction angle of the incident linear polarization $\gamma_i = 0°$. **c**, **g** and **k** are the target edge images of three different patterns with $\gamma_i = 10°$. **d**, **h** and **l** are the target edge images of three different patterns with $\gamma_i = 40°$.

In order to demonstrate that the differentiator can achieve broadband optical fully differential operation, we also used green laser beam (Coherent) with 532nm wavelength to conduct imaging experiments of edge detection. The experimental results are shown in Fig. 6. And experimental steps are the same as described in Fig. 5, the experimental results basically coincide with the results of He-Ne (632.8nm) laser beam. It proves that the differentiator can generate spatial differentiation within a certain bandwidth and can be applied to broadband edge detection.

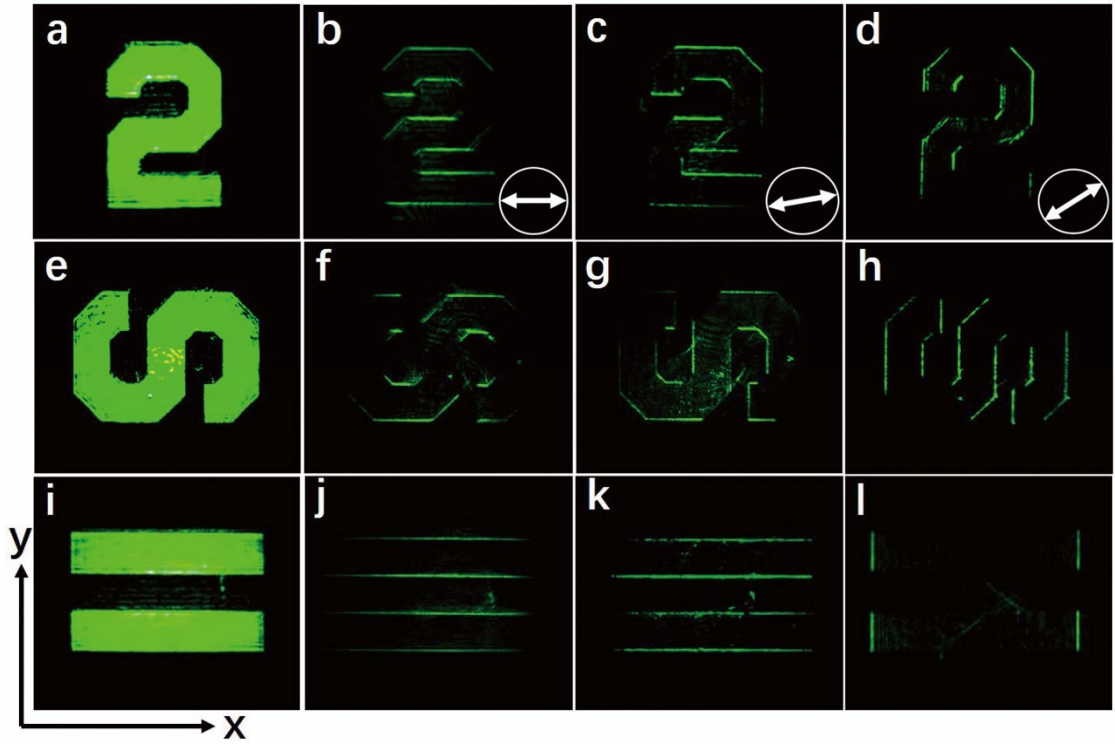

**Fig. 6 Edge detection for green laser (λ = 532 nm) with an incident angle $\theta_i = 30°$. a**, **e** and **i** are different incident images of number 2, the letter S and horizontal stripes, respectively. The white double arrow indicates the optical direction ($\gamma_i$) of GLP1. **b**, **f** and **j** are the target edge images of three different patterns with a direction angle of the incident linear polarization $\gamma_i = 0°$. **c**, **g** and **k** are the target edge images of three different patterns with $\gamma_i = 10°$. **d**, **h** and **l** are the target edge images of three different patterns with $\gamma_i = 40°$.

The proposed edge-detection mechanism can be briefly explained as follows: an object is illuminated by a linear polarized light and reflected at the air-glass interface. The left- and right-handed photons with the opposite spin angular momentum acquire opposite spatial shifts due to the spin-orbit interaction of light, which manifests spin-dependent images with a tiny shift at the image plane. The overlapped two spin components recombined to linear polarization thus will be eliminated by the analyzer, leaving out only the edge information available for detection.

## Discussion

In this paper, we have realized an optical fully differential operation based on the spin-orbit interaction of light at an optical interface. This differentiator can calculate the spatial difference of the input wavefunction in the process of light reflection, has the advantages of superspeed and low power consumption. It has the potential to promote optical computing instead of traditional computers based on electronic devices.

On the other hand, we demonstrate the application of broadband optical fully differentiator in edge detection, due to the purely geometric nature of spin-orbit interaction of light. Combine at least two edge detection images in orthogonal directions, which is enough to get a complete target edge image. This scheme can process the whole image in one shot, which is of great significance for high throughput real-time image processing. We believe that this optical fully differentiator will have a great potential in the field of microimaging. It should be noted that our scheme also works for single photons in quantum formalism, although it is demonstrated with classic light.

**Materials and methods**

The prism is mounted on the rotary table and the angle of incidence is controlled by the rotating prism. We measure the intensity of incident light and reflected light with a camera beam quality analyzer (BC106N-VIS from Thorlabs), compared with the scanning slit beam quality analyzer, the camera beam quality analyzer can capture a more detailed beam profile and provide a two-dimensional analysis of the true beam power density distribution. Considering the influence of lens and polarizer on the beam intensity, the spatial spectral transfer function is obtained by processing the data. In the experiment of edge detection, in order to completely cover the target image with the cross section of laser beam, we choose a beam expander of 5 times.

**Acknowledgements**

This research was supported by the National Natural Science Foundation of China (Grant No. 61835004).

**Author contributions**

H. L. conceived of the original concept. H. L. and S. W. supervised the project. S. H. and J. Z. designed the research. The experiment was designed and performed by S. H. and J. Z.. The data was analyzed by S. H. J. Z., S. C., and W. S.. All authors discussed the results and co-wrote the paper.

**Additional information**

Supplementary Information is available in the online version of the paper. Reprints and permissions information is available online. Correspondence and requests for materials should be addressed to L.H.

**Competing financial interests**

The authors declare no competing financial interests.

# Broadband Optical Fully Differential Operation Based on the Spin-orbit Interaction of Light

Shanshan He[1], Junxiao Zhou[1], Shizhen Chen[1], Weixing Shu[2], Hailu Luo[1*] and Shuangchun Wen[2]

[1]Laboratory for Spin Photonics, School of Physics and Electronics, Hunan University, Changsha 410082, China

[2]Key Laboratory for Micro-/Nano-Optoelectronic Devices of Ministry of Education, School of Physics and Electronics, Hunan University, Changsha 410082, China

*Hailu Luo hailuluo@hnu.edu.cn

**Supplementary information**

**Section 1. Theoretical calculation of optical fully differentiational operation**

The origin of spin–orbit interaction of light upon reflection at an optical interface is attributed to the transverse nature of the photonic polarization: The polarizations of angular spectrum components experience different rotations in order to satisfy the transversality[S1-S3]. We start with a horizontally $|H\rangle$ or vertically $|V\rangle$ polarized incident wave-packet. After reflection, the polarization states evolve as [S4]

$$|H(k_i)\rangle \to r_p[|H(k_r)\rangle + k_{rx}\Delta_H|H(k_r)\rangle + k_{ry}\delta_H|V(k_r)\rangle], \qquad (S1)$$

$$|V(k_i)\rangle \to r_s[|V(k_r)\rangle + k_{rx}\Delta_V|H(k_r)\rangle - k_{ry}\delta_V|H(k_r)\rangle], \qquad (S2)$$

where

$$\delta_H = \frac{(r_p+r_s)\cot\theta_i}{k_0 r_p}, \qquad (S3)$$

$$\delta_V = \frac{(r_p+r_s)\cot\theta_i}{k_0 r_s}, \qquad (S4)$$

$$\Delta_H = \frac{\partial \ln r_p}{k_0 \partial \theta_i}, \qquad (S5)$$

$$\Delta_V = \frac{\partial \ln r_s}{k_0 \partial \theta_i}. \qquad (S6)$$

In the spin space $|H\rangle = \frac{1}{\sqrt{2}}(|+\rangle + |-\rangle)$, $|V\rangle = \frac{i}{\sqrt{2}}(|-\rangle - |+\rangle)$, we have

$$|H(k_i)\rangle \to \frac{r_p}{\sqrt{2}}[(1 + k_{rx}\Delta_H - ik_{ry}\delta_H)|+\rangle + (1 + k_{rx}\Delta_H + ik_{ry}\delta_H)|+\rangle], \qquad (S7)$$

$$|V(k_i)\rangle \to -\frac{ir_s}{\sqrt{2}}[(1 + k_{rx}\Delta_V - ik_{ry}\delta_V)|+\rangle - (1 + k_{rx}\Delta_H + ik_{ry}\delta_H)|+\rangle]. \qquad (S8)$$

For an arbitrary incident linear polarization state,

$$|\psi_i\rangle = \cos\gamma_i |H\rangle + \sin\gamma_i |V\rangle, \tag{S9}$$

the polarization state of reflected wave packet can be written as

$$|\psi_r\rangle \approx \frac{1}{\sqrt{2}}(\cos\gamma_i\, r_p - i\sin\gamma_i\, r_s)[1 + ik_{rx}\Delta x + ik_{ry}\Delta y]|+\rangle + \frac{1}{\sqrt{2}}(\cos\gamma_i\, r_p + i\sin\gamma_i\, r_s)[1 - ik_{rx}\Delta x - ik_{ry}\Delta y]|-\rangle, \tag{S10}$$

where

$$\Delta x = \frac{r_p r_s \tan\gamma_i}{r_p^2 + r_s^2 \tan^2\gamma_i}(\Delta_H - \Delta_V), \tag{S11}$$

$$\Delta y = \frac{r_p^2 \delta_H + r_s^2 \tan^2\gamma_i\, \delta_V}{r_p^2 + r_s^2 \tan^2\gamma_i}. \tag{S12}$$

For weak spin-orbit interaction of light $\Delta x, \Delta y \ll w_0$, the wavefunction can be written as

$$|\psi_r\rangle \approx \frac{|A|}{\sqrt{2}}\{\exp(+ik_{rx}\Delta x + ik_{ry}\Delta y)\exp(-i\gamma_r)|+\rangle + \exp(-ik_{rx}\Delta x - ik_{ry}\Delta y)\exp(+i\gamma_r)|-\rangle\}, \tag{S13}$$

where $|A| = \sqrt{(\cos\gamma_i\, r_p)^2 + (\sin\gamma_i\, r_s)^2}$, $\gamma_r = \arctan\left(\frac{r_s}{r_p}\tan\gamma_i\right)$. Here, we have introduced the approximations: $\exp i\sigma_3(k_{rx}\Delta x + k_{ry}\Delta y) \approx 1 + i\sigma_3(k_{rx}\Delta x + k_{ry}\Delta y)$.

By taking the Fourier transform of the angular spectrum of reflected light field, we get the reflected wavefunction in position space:

$$\varphi_r(x, y) = \iint \varphi_{in}(k_{ix}, k_{iy}) \exp[i(k_{rx}x + k_{ry}y)] dk_{rx}\, dk_{ry}. \tag{S14}$$

Based on the boundary conditions $k_{iy} = k_{ry}$, $k_{ix} = -k_{rx}$, we have

$$|\varphi_r(x, y)\rangle = \frac{1}{\sqrt{2}}|A|\exp(-i\gamma_r)\varphi_{in}(x, y)[(x + \Delta x), (y + \Delta y)]|+\rangle + \frac{1}{\sqrt{2}}|A|\exp(+i\gamma_r)\varphi_{in}(x, y)[(x - \Delta x), (y - \Delta y)]|-\rangle. \tag{S15}$$

Then the reflected wave packet passes through second polarizer whose polarization axis is chosen as $\gamma_{p2}$, and the Jones matrix can be written as

$$J = \begin{bmatrix} \cos^2\gamma_{p2} & \sin\gamma_{p2}\cos\gamma_{p2} \\ \sin\gamma_{p2}\cos\gamma_{p2} & \sin^2\gamma_{p2} \end{bmatrix}. \tag{S16}$$

Hence, the final electrical field through the whole differentiator system can be obtained

as

$$|\varphi_{out}(x,y)\rangle = -\frac{i}{2}|A|[\varphi_{in}[(x+\Delta x),(y+\Delta y)] - \varphi_{in}[(x-\Delta x),(y-\Delta y)]](\sin\gamma_r|H\rangle - \cos\gamma_r|V\rangle), \quad (S17)$$

where $\gamma_{p2} = \frac{\pi}{2} + \gamma_r$. If the spin-dependent shifts $\Delta x$, $\Delta y$ are much smaller than the width of incident wavepacket, $|\varphi_{out}(x,y)\rangle$ is approximately proportional to the first-order spatial differentiation of the input $|\varphi_{in}(x,y)\rangle$:

$$|\varphi_{out}(x,y)\rangle = -i|A|\left[\Delta x \frac{\partial \varphi_{in}(x,y)}{\partial x} + \Delta y \frac{\partial \varphi_{in}(x,y)}{\partial y}\right](\sin\gamma_r|H\rangle - \cos\gamma_r|V\rangle). \quad (S18)$$

**Section 2. The spatial full differentiation**

In general, the spin-dependent shifts are much smaller than the profile of input wave function due to the weak spin-orbit interaction of light. Therefore, $\varphi_{out}(x,y)$ can be approximately written as the spatial full differentiation of the input $\varphi_{in}(x,y)$:

$$\varphi_{out}(x,y) \simeq \Delta x \frac{\partial \varphi_{in}(x,y)}{\partial x} + \Delta y \frac{\partial \varphi_{in}(x,y)}{\partial y}. \quad (S19)$$

Here, $\Delta x$ and $\Delta y$ are increments which can be modulated by the incident polarization $\gamma_i$. The splitting direction angle is obtained as $\alpha = \arctan(\Delta y/\Delta x)$ as shown in Figure S1. To generate splitting light beams at any desirable direction and optimal edge images, the incident angle is chosen as $\theta_i = 30°$.

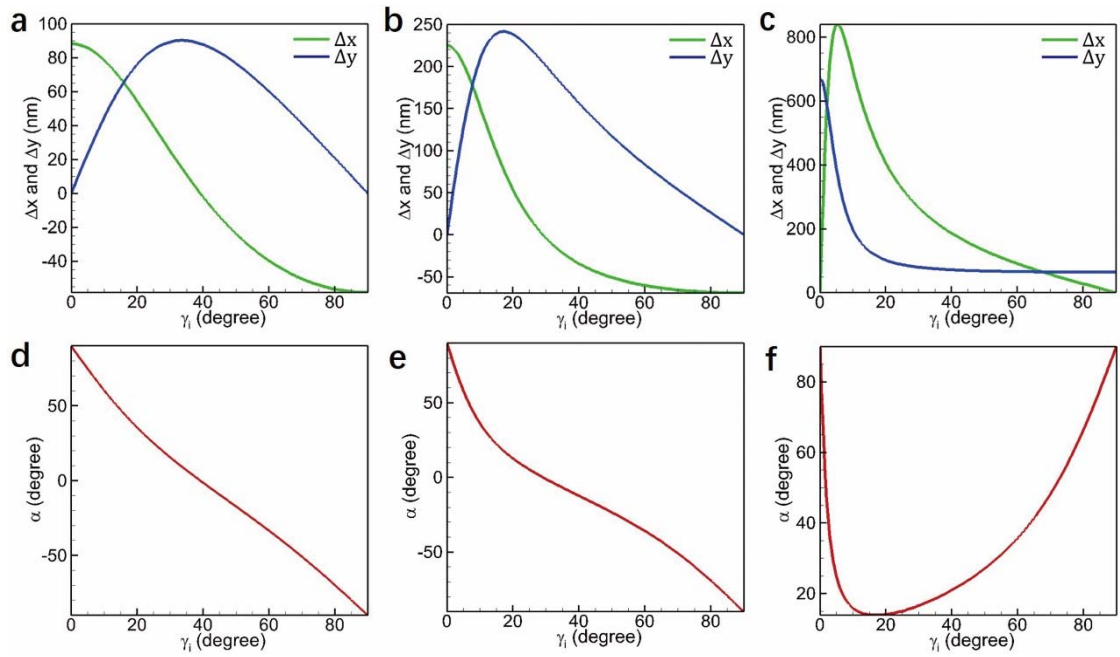

**Figure S1. Spatial differentiation varies with incident polarization angle $\gamma_i$. a-c** The spin-dependent shifts as a function of polarization angle of incident beam for different incident angles $\theta_i = 30°$, $45°$, and $60°$. **d-f** The corresponding angle of splitting direction.

**Supplementary References**